\documentclass[sigconf,natbib=true]{acmart}
\usepackage{multirow}
\usepackage{subcaption}
\usepackage{color}
\usepackage{enumitem}

\usepackage{tcolorbox}

\AtBeginDocument{%
  \providecommand\BibTeX{{%
    \normalfont B\kern-0.5em{\scshape i\kern-0.25em b}\kern-0.8em\TeX}}}

\setcopyright{acmlicensed}
\copyrightyear{2018}
\acmYear{2018}
\acmDOI{XXXXXXX.XXXXXXX}

\acmConference[Conference acronym 'XX]{Make sure to enter the correct
  conference title from your rights confirmation emai}{June 03--05,
  2018}{Woodstock, NY}
\acmBooktitle{Woodstock '18: ACM Symposium on Neural Gaze Detection,
 June 03--05, 2018, Woodstock, NY} 
\acmISBN{978-1-4503-XXXX-X/18/06}

\setlist[itemize]{leftmargin=*}

\begin{document}

\title{Fusion and Alignment Enhancement with Large Language Models for Tail-item Sequential Recommendation}

\author{Zhifu Wei}
\affiliation{%
  \institution{Northeastern University}
  \city{Shenyang}
  \country{China}
}
\email{weizf@mails.neu.edu.cn}

\author{Yizhou Dang}
\affiliation{%
  \institution{Northeastern University}
  \city{Shenyang}
  \country{China}
}
\email{dangyz@mails.neu.edu.cn}

\author{Guibing Guo}
\authornote{Corresponding author.}
\affiliation{%
  \institution{Northeastern University}
  \city{Shenyang}
  \country{China}
}
\email{guogb@swc.neu.edu.cn}

\author{Chuang Zhao}
\affiliation{%
  \institution{Tianjin University}
  \city{Tianjin}
  \country{China}
}
\email{zhaochuang@tju.edu.cn}

\author{Zhu Sun}
\authornotemark[1]
\affiliation{%
  \institution{Singapore University of Technology and Design}
  \city{Singapore}
  \country{Singapore}
}
\email{sunzhuntu@gmail.com}

\renewcommand{\shortauthors}{Zhifu and Yizhou, et al.}

\begin{abstract}

Sequential Recommendation (SR) learns user preferences from their historical interaction sequences and provides personalized suggestions. In real-world scenarios, most items exhibit sparse interactions, known as the tail-item problem. This issue limits the model's ability to accurately capture item transition patterns. To tackle this, large language models (LLMs) offer a promising solution by capturing semantic relationships between items. Despite previous efforts to leverage LLM-derived embeddings for enriching tail items, they still face the following limitations: 1) They struggle to effectively fuse collaborative signals with semantic knowledge, leading to suboptimal item embedding quality. 2) Existing methods overlook the structural inconsistency between the ID and LLM embedding spaces, causing conflicting signals that degrade recommendation accuracy. In this work, we propose a \textbf{F}usion and \textbf{A}lignment \textbf{E}nhancement framework with LLMs for Tail-item Sequential \textbf{Rec}ommendation (\textbf{FAERec}), which improves item representations by generating coherently-fused and structurally consistent embeddings. For the information fusion challenge, we design an adaptive gating mechanism that dynamically fuses ID and LLM embeddings. Then, we propose a dual-level alignment approach to mitigate structural inconsistency. The item-level alignment establishes correspondences between ID and LLM embeddings of the same item through contrastive learning, while the feature-level alignment constrains the correlation patterns between corresponding dimensions across the two embedding spaces. Furthermore, the weights of the two alignments are adjusted by a curriculum learning scheduler to avoid premature optimization of the complex feature-level objective. Extensive experiments across three widely used datasets with multiple representative SR backbones demonstrate the effectiveness and generalizability of our framework. The codes are provided at \url{https://github.com/ZhifuWei/FAERec}.

\end{abstract}





\maketitle

\section{Introduction}
Sequential recommendation aims to predict the next items of interest to users based on their historical interaction behavior and has been widely applied across various real-world scenarios such as e-commerce, streaming media, and social networks \cite{tang2018personalized,hidasi2015session,kang2018self}. 
Learning high-quality item embeddings is essential for models to effectively capture  item transition patterns and users’ behaviors \cite{hu2025alphafuse,dang2024augmenting,dang2026exploring}.
Early research learned ID embeddings for each item, capturing collaborative signals through user-item interaction data \cite{jang2020cities,li2017neural}.

\begin{figure}[!t]
  \centering
  \includegraphics[width=0.47\textwidth]{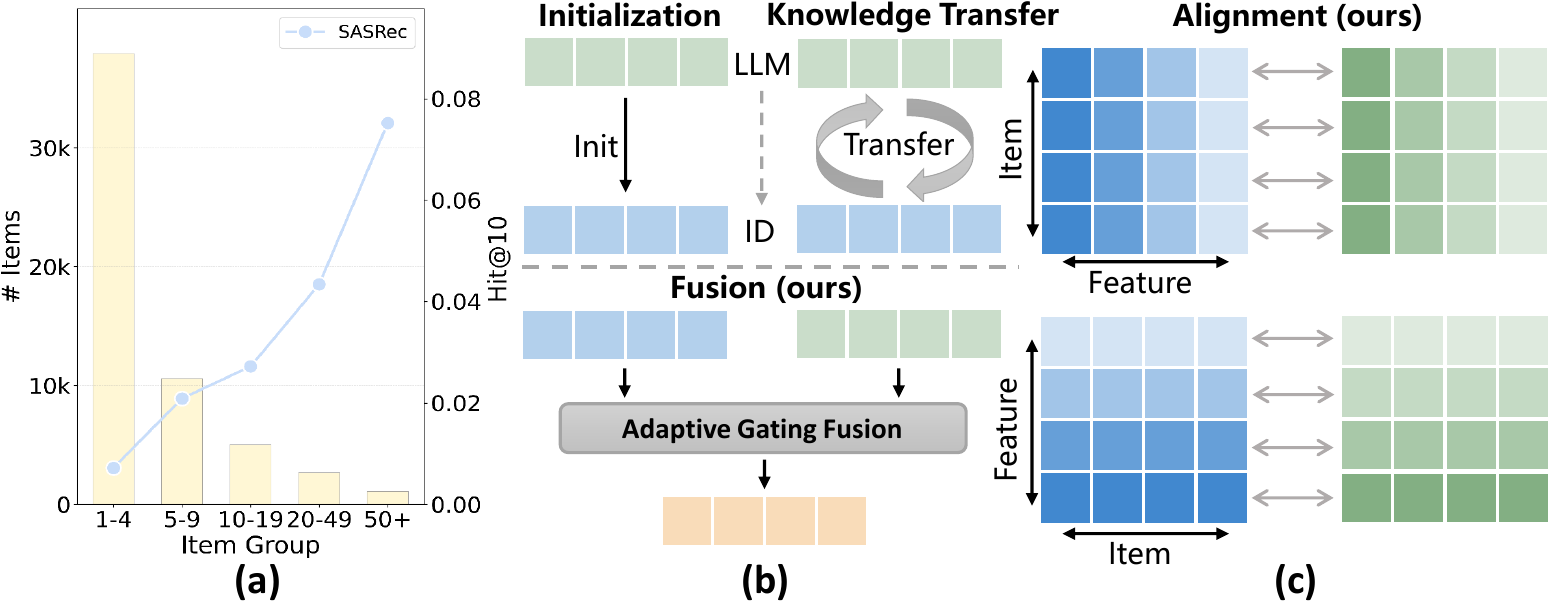}
  \vspace{-1em}
  \caption{Visualization of the long-tail item problem (a) and comparison between the existing LLM-based sequential recommendation methods and our FAERec (b, c).}
  \label{fig:example}
  \vspace{-1.5em}
\end{figure}

Despite significant advances in sequential recommendation, the tail-item problem remains a critical challenge, undermining item embedding quality and user experience \cite{li2025reembedding,klimashevskaia2024survey,dang2026tail}. As shown in Figure \ref {fig:example} (a), we trained SASRec \cite{kang2018self}, a representative sequential recommendation model, on the Amazon Beauty dataset and display its performance across items grouped by interaction frequency. The histogram reveals that most items have fewer
than 5 records, while the corresponding line graph indicates their relatively low performance. This stems from an inherent limitation of ID-based models, which rely solely on collaborative signals from interaction data. For tail items with sparse interactions, such models fail to provide adequate information for effective representation learning \cite{jang2020cities}. In contrast, abundant interactions enable head items to leverage rich collaborative patterns, resulting in significantly better performance.


To tackle the tail-item challenge, recent advances in large language models (LLMs) offer new opportunities by capturing semantic item relationships from textual attributes \cite{harte2023leveraging,liu2025llmemb,liu2025bridge,liu2025large}. Motivated by this, several semantic-guided learning strategies have been proposed to enhance ID embeddings \cite{liu2024llm,liu2025llmemb,qu2024elephant,zhang2024recdcl,hu2025alphafuse}. 1) \textbf{Semantic initialization} \cite{harte2023leveraging,qu2024elephant,wang2025pre} utilizes language embeddings to provide effective initialization for the ID embedding layer, gradually evolving from semantic space to collaborative space.
2) \textbf{Knowledge transfer} \cite{ren2024representation,liu2025llmemb} maximizes mutual information between ID and LLM embeddings through cross-view contrastive learning.

Despite their effectiveness, these methods face two critical challenges:  (i) \textbf{Insufficient Semantic Leveraging.} As shown in Figure \ref{fig:example} (b), semantic initialization and knowledge transfer strategies use LLM embeddings solely as auxiliary signals for learning ID embeddings, without explicitly retaining LLM embeddings in the final item embeddings, leading to the suboptimal item embedding quality. (ii) \textbf{Inconsistent Space Structures.} While achieving cross-view knowledge transfer, existing methods overlook the structural alignment between the ID and LLM embedding spaces. Specifically, the embedding space structure is determined by how each dimension encodes item information and how dimensions correlate with each other \cite{zbontar2021barlow,ermolov2021whitening,liuimproving}. Since ID embeddings capture collaborative patterns while LLM embeddings encode semantic textual attributes, the two spaces organize information through inconsistent dimensional structures \cite{liu2024alignrec}. This results in inconsistent item neighborhoods across the two embedding spaces: Items close in collaborative space may be distant in semantic space (\emph{due to space limitations, this is verified  in Section~\ref{subsec:visualization}}). Such structural inconsistency introduces conflicting signals when leveraging LLM semantic knowledge, thereby degrading recommendation accuracy.

In this paper, we propose a \textbf{F}usion and \textbf{A}lignment \textbf{E}nhancement framework with LLMs for Tail-item Sequential \textbf{Rec}ommendation (\textbf{FAERec}). Our core idea is to improve semantic utilization and structural consistency by effectively fusing and aligning ID and LLM embeddings. Specifically, we first derive semantic embeddings by encoding item texts with LLMs, which can be cached to avoid extra inference costs.
To further leverage semantic information, we perform dimension-wise weighted fusion of ID and LLM embeddings, allowing the model to capture fine-grained complementary patterns between collaborative and semantic signals. The fusion weights are adaptively determined by a gating mechanism based on the characteristics of each item's embeddings. For structural inconsistency, as shown in Figure \ref{fig:example} (c), we design a dual-level alignment approach.
The item-level alignment utilizes contrastive learning to maximize the similarity between ID and LLM embeddings of the same item while minimizing it for different items, establishing point-to-point correspondences across the two embeddings. 
At the feature level, inspired by the Barlow Twins \cite{zbontar2021barlow}, we first standardize the ID and LLM embeddings separately and compute their cross-correlation matrix. Furthermore, we constrain this matrix to align corresponding dimensions while decoupling different dimensions, achieving structural consistency across embedding spaces.
To prevent premature convergence to the complex feature-level objective, we introduce a curriculum learning scheduler based on cosine annealing, balancing the two alignment objectives.
Extensive experiments on various datasets
and backbones demonstrate that our proposed method achieves significant improvements.

In summary, our work makes the following contributions:
\begin{itemize}
    \item We highlight insufficient semantic leveraging and inconsistent space structures of existing LLM-based SR methods and propose a \textbf{F}usion and \textbf{A}lignment \textbf{E}nhancement framework with LLMs for Sequential \textbf{Rec}ommendation, which can improve the quality of item embeddings and alleviate the tail-item challenge.
    
    \item We design two key modules. Adaptive gating fusion effectively integrates collaborative signals and semantic knowledge. Dual-level alignment establishes item-level correspondences between ID and LLM embeddings and feature-level structural consistency across the two embedding spaces, guided by curriculum learning.

    \item We conduct comprehensive experiments on real-world datasets with representative backbone SR models and different baselines to demonstrate the effectiveness and generalizability of FAERec.
\end{itemize}

\section{RELATED WORK}
\subsection{General Sequential Recommendation}
Sequential Recommendation (SR) aims to predict users' next item of interest based on their interaction sequences \cite{zhao2023sequential,zhao2023cross,dang2024data}. Early research leveraged Markov Chains \cite{he2016fusing} to extract user preference, focusing on correlations between the next item and recent interactions. Later, deep learning architectures such as CNNs \cite{yuan2019simple} and RNNs \cite{2022cmnrec} were adopted: GRU4Rec \cite{hidasi2015session} used gated recurrent units to model interest evolution, while Caser \cite{tang2018personalized} projected items to latent spaces and learned patterns via convolution operations. With the advancement of self-attention \cite{vaswani2017attention,zhang2024cf}, Transformer-based SR models gained prominence. SASRec \cite{kang2018self} adopted the multi-head attention mechanism to characterize transition correlations effectively. BERT4Rec \cite{sun2019bert4rec} introduced a two-directional attention encoder trained with a cloze-style prediction task. Departing from Transformer-based methods, research showed that filtering mechanisms effectively mitigate noise \cite{zhou2022filter}, which motivated the proposal of an all-MLP framework with trainable filters. To efficiently learn sequential transitions, LRURec \cite{yue2024linear} introduced matrix diagonalization-based linear recurrence, complemented by a recursive parallelization scheme that improves training speed.

\subsection{Long-tail Sequential Recommendation}
To alleviate the tail-item problem, early research focuses on leveraging collaborative signals from head items with abundant interactions to enrich tail-item representations \cite{jang2020cities,kim2023melt}. As a representative work, CITIES \cite{jang2020cities} leverages multiple contextual items from the head to train an embedding inference function, improving the embedding quality of tail items. Inspired by this idea, MELT \cite{kim2023melt} designs a mutually reinforcing dual-branch framework to enhance performance for both tail users and tail items by transferring knowledge from head users and items, respectively. MASR \cite{hu2022memory} proposes centroid-wise and cluster-wise memory banks to store historical interaction patterns and retrieve similar sequences for tail items. LOAM \cite{yang2023loam} generates augmented interaction sequences that end with tail items through global context integration, and leverages cross-session mixup to synthesize new interactions to enrich tail item representations.
Despite their effectiveness, these approaches fundamentally rely on collaborative signals extracted from interaction data. Tail items with highly sparse interactions fail to provide sufficient contextual information for these methods to learn effective representations, resulting in suboptimal item embeddings.

\subsection{LLMs for Recommendation} Large language models have attracted widespread attention due to their powerful abilities in semantic understanding \cite{zhao2023survey,liu2025fedadamw,zhi2025reinventing,zhao2025grounded}.
Recent research has explored integrating LLMs into recommender systems, which can be categorized into two primary approaches \cite{liu2025llmemb,hu2025alphafuse,qu2024elephant,wu2024exploring,wang2024llm4msr,li2024calrec}.
The first approach directly employs LLMs for recommendation tasks. For example, TALLRec \cite{bao2023tallrec} uses instruction tuning to enable LLMs to generate recommendations. ChatRec \cite{gao2023chat} designs prompt templates to activate LLM capabilities. The second approach leverages LLM embeddings to enhance traditional models. Semantic initialization methods, such as LLM2Bert4Rec \cite{harte2023leveraging} and SAID \cite{hu2024enhancing}, utilize LLM embeddings to initialize ID embeddings, which subsequently adapt to collaborative patterns during training. RLMRec \cite{ren2024representation} proposes knowledge transfer through mutual information maximization, establishing bidirectional mappings between language and behavior spaces. LLM-ESR \cite{liu2024llm} introduces dual-view modeling and retrieval-augmented distillation to improve representations for both tail users and items in sequential recommendation. Unlike the above efforts, our proposed FAERec is tailored from both the fusion and alignment perspectives, furthermore advancing LLM-driven item representation learning for tail items. 

\section{PRELIMINARIES}

\textbf{Problem Definition.}Suppose we have a set of users $\mathcal{U}$ and a set of items $\mathcal{V}$. Each user $u \in \mathcal{U}$ corresponds to a chronologically ordered sequence of interacted items $s_{u} = [v_{1}, ..., v_{j}, ..., v_{|s_{u}|}]$, where $v_{j} \in \mathcal{V}$ denotes the item $u$ interacted with at time step $j$, and $|s_{u}|$ is the sequence length. Given a user's interaction item sequence $s_{u}$, the sequence recommendation objective is to accurately predict the item $v^{*}$ most likely to be interacted with by user $u$ at time step $|s_{u}|+1$. Its mathematical expression is:
\begin{equation}
\underset{v^{*} \in \mathcal{V}}{\text{arg max}}           P\left(v_{|s_{u}|+1}=v^{*} \mid s_{u}\right)
\end{equation}
\textbf{Head/Tail Splitting.}
Referring to existing research on long-tail item sequential recommendations \cite{liu2024llm,kim2023melt}, we categorize items into tail and head groups. Let $p_v$ denote the popularity of item $v$. We sort items in descending order based on $p_v$. Subsequently, following the Pareto principle \cite{box1986analysis}, we select the top 20\% of items as head items ($\mathcal{V}_{\text{head}}$), with the remaining items constituting the tail items, i.e., $\mathcal{V}_{\text{tail}} = \mathcal{V} \setminus \mathcal{V}_{\text{head}}$ \cite{liu2024llm}. To mitigate the tail-item challenge, we aim to enhance recommendation performance specifically for $\mathcal{V}_{\text{tail}}$.

\begin{figure*}[!t]
  \centering
  \includegraphics[scale=0.557]{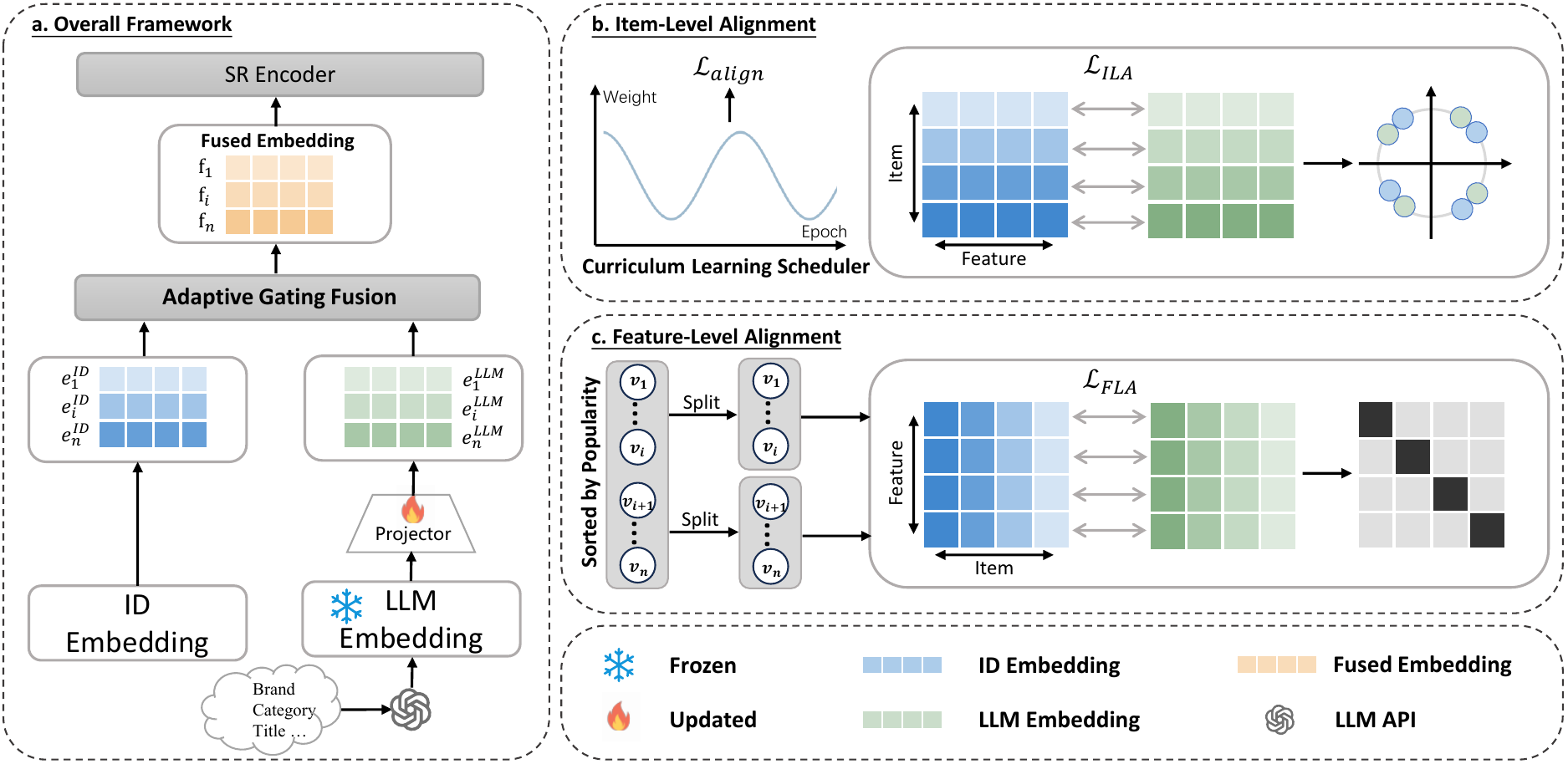}
  \vspace{-1em}
  \caption{The overall framework of our proposed FAERec.}
  \label{fig:framework}
  \vspace{-1em}
\end{figure*}

\section{OUR METHOD}
In this section, we present our proposed FAERec. The Overview framework is illustrated in Figure \ref{fig:framework}. Our approach consists of two key modules: adaptive gating fusion and dual-level alignment. We then introduce the training and inference processes of FAERec. Finally, we discuss our FAERec and existing methods.

\subsection{Adaptive Gating Fusion}
Traditional SR models use randomly initialized ID embeddings to learn collaborative patterns, achieving strong performance on popular items but struggling with tail items due to sparse interactions \cite{li2025reembedding,kim2023melt}. To address this, we enhance item embeddings by fusing collaborative signals from ID embeddings with semantic knowledge from LLM embeddings. Besides, we introduce a dimension-wise weighting strategy to combine the benefits of both better.

\vspace{0.3em}

\noindent\textbf{ID Embedding Layer.} For each item $v_i \in \mathcal{V}$, we maintain a trainable embedding vector $\mathbf{e}^{ID}_i \in \mathbb{R}^d$, forming the ID embedding matrix $\mathbf{E}^{ID} \in \mathbb{R}^{|\mathcal{V}| \times d}$. These embeddings are updated during model training via backpropagation to capture underlying item transition patterns and collaborative signals. Given an original user sequence $s_u$, the embedding Look-up operation will be applied for $E^{ID}$ to obtain an item embedding sequence, i.e, $E^{ID}_u=[\mathbf{e}^{ID}_1, \mathbf{e}^{ID}_2, \ldots, \mathbf{e}^{ID}_{|s_{u}|}]$.

\vspace{0.3em}

\noindent\textbf{LLM Embedding Layer.} Typically, the attributes and descriptions of items contain rich semantic information. To leverage the text comprehension capabilities of LLMs, we organize item attributes (e.g., title, category and description) into natural language prompts and encode them using a pre-trained LLM to obtain LLM embeddings \cite{ren2024representation}. Specifically, we construct the following item prompts:
\definecolor{natblue}{RGB}{70,130,180}        
\definecolor{paleblue}{RGB}{230,240,250}       
\begin{tcolorbox}[
    colback=white,
    colframe=natblue,
    colbacktitle=paleblue,
    coltitle=natblue,
    title={Item Prompt Template (Beauty)},
    fonttitle=\bfseries,
    arc=2mm,
    boxrule=0.8pt
]
The Beauty item has the following attributes:\\
name is \underline{<TITLE>}; brand is \underline{<BRAND>}; rating is \underline{<DATE>};\\
price is \underline{<PRICE>}.

The item has following features: \underline{<CATEGORIES>}.

The item has following descriptions: \underline{<DESCRIPTION>}.
\end{tcolorbox}
By substituting the template fields with corresponding attribute values, item prompts are input to LLMs to extract  rich semantic embeddings. In practice, LLM embeddings can be obtained through open-source models like LLaMA \cite{touvron2023llama} or commercial APIs such as OpenAI's text-embedding-ada-002 \footnote{\url{https://platform.openai.com/}}. We employ the latter to ensure high-quality and stable embeddings. Let $\mathbf{E}^{LLM} \in \mathbb{R}^{d_{llm}}$ denote the LLM embeddings of all items, where $d_{llm}$ is the embedding dimension. All item LLM embeddings are generated offline in advance  and cached before training, thereby avoiding the computational overhead of online inference.
Besides, since the dimension $d_{LLM}$ of original LLM embeddings (e.g., 1536 dimensions) is significantly higher than the hidden dimension $d$ of the recommendation model (typically 64 or 128), we designed a lightweight projection network to perform dimension adaptation\cite{hu2024enhancing}.  Specifically, we obtain the original LLM embedding $\mathbf{e}_{ori}^{LLM}$ from the LLM API, and project it to match the ID embedding dimension:
\begin{equation}
\mathbf{e}_i^{LLM} = f_\theta(\mathbf{e}_{ori}^{LLM})
\end{equation}
where $f_\theta(\cdot)$ is implemented via two layers of MLP. Then, we obtain the LLM embeddings corresponding to user $u$'s interaction sequence, denoted as $\mathbf{E}^{LLM}_u=[\mathbf{e}^{LLM}_1, \mathbf{e}^{LLM}_2, \ldots, \mathbf{e}^{LLM}_{|s_u|}]$. The projection network parameters are trainable, while the original LLM embeddings remain frozen throughout training .
Besides, since LLM embeddings are obtained from pretrained models while ID embeddings are learned from scratch, directly combining them may lead to an imbalance in the optimization \cite{liu2024llm,amos2023never}. To mitigate this, we use Principal Component Analysis (PCA) to reduce LLM embeddings to $d$ dimensions and use them to initialize ID embeddings.

\vspace{0.3em}

\noindent\textbf{Dimension-wise Weighted Fusion.} While ID embeddings effectively capture collaborative patterns for 
head items with abundant interactions, LLM embeddings provide more reliable semantic information for tail items with sparse data \cite{wang2025empowering,liu2025llmemb}. To leverage these complementary strengths, we apply dimension-wise weighting to integrate the two embeddings, allowing the model to balance collaborative and semantic information at each dimension.

For the item $v_i$, we concatenate its ID embedding $\mathbf{e}^{ID}_i$ and LLM embedding $\mathbf{e}^{LLM}_i$ and feed them into the gating network:
\begin{equation}
  \mathbf{g}_i = \sigma\left(\mathbf{W}_2 (\mathbf{W}_1 [\mathbf{e}^{ID}_i \oplus \mathbf{e}^{LLM}_i] + \mathbf{b}_1) + \mathbf{b}_2\right)  
\end{equation}
where $\mathbf{W}_1 \in \mathbb{R}^{2d \times 2d}$, $\mathbf{W}_2 \in \mathbb{R}^{d \times 2d}$, $\mathbf{b}_1 \in \mathbb{R}^{2d}$, and $\mathbf{b}_2 \in \mathbb{R}^{d}$ are the weight matrices and biases of the gating network. $\sigma(\cdot)$ denotes the Sigmoid function, and the output gating vector $\mathbf{g}_i \in \mathbb{R}^d$ takes 
values in $[0, 1]$ for each dimension. The final fused embedding is obtained via dimension-wise weighted interpolation:
\begin{equation}
    \mathbf{f}_i = \mathbf{g}_i \odot \mathbf{e}^{ID}_i + (\mathbf{1} - \mathbf{g}_i) \odot \mathbf{e}^{LLM}_i
\end{equation}
After applying this fusion to all items interacted by user $u$, we obtain the fused item embedding sequence $\mathbf{F}_u = [\mathbf{f}_1, \ldots, \mathbf{f}_{|s_u|}]$. The sequence is then fed into the SR encoder to produce the output representation of the sequence $H_u$.

\subsection{Dual-Level Alignment}
Existing LLM-based methods typically enhance ID embeddings directly leveraging LLM embeddings, without considering whether the structure of the two embedding spaces is aligned \cite{liu2024llm,hu2025alphafuse}. This structural 
mismatch introduces conflicting signals when integrating collaborative and semantic information, undermining the effectiveness of LLM enhancement. To tackle this issue, we propose a dual-level alignment approach, which progressively aligns the two embedding spaces from point-to-point correspondences at the item level to structural consistency at the feature level.

\vspace{0.3em}

\noindent\textbf{Item-Level Alignment.} The first level of alignment aims to establish point-to-point correspondences between ID and LLM embeddings of the same item across the two spaces. We introduce contrastive learning \cite{xie2022contrastive,chen2020simple} to achieve this. Specifically, for a training batch of $K$ items, the item-level alignment loss is defined as:
\begin{equation}
    \mathcal{L}_{ILA} = -\frac{1}{K} \sum_{i=1}^{K} \log \frac{\exp(\text{sim}(\mathbf{e}^{ID}_i, \mathbf{e}^{LLM}_i)/\tau)}{\sum_{j=1}^ {K} \exp(\text{sim}(\mathbf{e}^{ID}_i, \mathbf{e}^{LLM}_j)/\tau)}
\end{equation}
where $\text{sim}(\cdot, \cdot)$ denotes cosine similarity and $\tau$ is the temperature hyperparameter of contrastive learning. This loss function treats ID-LLM embedding pairs for the same item as positive samples, and pairs from different items as negative samples.

\noindent\textbf{Feature-Level Alignment.} Recent studies demonstrate that constraining the correlation structure of feature dimensions can effectively improve embedding quality \cite{zhang2024recdcl,ermolov2021whitening}. Inspired by this, we adopt the Barlow Twins \cite{zbontar2021barlow} to achieve alignment at the feature-level. Specifically, we construct a cross-view correlation matrix that enforces high correlation between corresponding dimensions while decoupling different dimensions, achieving structural consistency between the ID and LLM embedding spaces.

Unlike item-level alignment, feature-level alignment requires calculating statistics at the batch dimension. Highly popular items, due to their higher interaction frequency, dominate the normalization process, thereby weakening the alignment effect for tail items. Therefore, we use the median popularity $p_{\text{median}}$ of batch items as a threshold to partition the batch item set into high-popularity and low-popularity item sets. The median strategy balanced group 
sizes and prevents the model training from being dominated by either group. The process is formalized as follows:
\begin{equation}
   \mathcal{V}^{B_h} = \{i \mid p_i \geq p_{median}\}, \quad \mathcal{V}^{B_l} = \{i \mid p_i < p_{median}\} 
\end{equation}
where $p_i$ denotes the popularity of item $i$. We illustrate the progress using $\mathcal{V}^{B_h}$ as an example, with $\mathcal{V}^{B_l}$ following the same process.

For the $S$ item samples in the high-popularity group, we first standardize the ID embeddings and LLM embeddings along the batch dimension separately. Specifically, for each embedding matrix $\mathbf{E} \in \mathbb{R}^{S \times d}$, its standardized form  is defined as follows:
\begin{equation}
    \tilde{\mathbf{e}}^{B_h}_{n,i} = \frac{\mathbf{e}^{B_h}_{n,i} - \mu^{B_h}_i}{\sigma^{B_h}_i}
\end{equation}
where $\mu^{B_h}_i$ and $\sigma^{B_h}_i$ respectively denote the mean and standard deviation of the $i$-th dimension of the embedding matrix $\mathbf{E}$. The normalized embedding matrices are denoted as $\tilde{\mathbf{E}}^{\text{ID}}, \tilde{\mathbf{E}}^{\text{LLM}} \in \mathbb{R}^{S \times d}$.

Based on the standardized embeddings, we conduct the cross-correlation matrix $\mathbf{C}^{B_h} \in \mathbb{R}^{d \times d}$, whose elements are defined as:
\begin{equation}
  \mathbf{C}^{B_h}_{ij} = \frac{\sum_{s=1}^{S} \tilde{\mathbf{e}}^{ID}_{s,i} \tilde{\mathbf{e}}^{LLM}_{s,j}}{\sqrt{\sum_{s=1}^{S} (\tilde{\mathbf{e}}^{ID}_{s,i})^2} \sqrt{\sum_{s=1}^{S} (\tilde{\mathbf{e}}^{LLM}_{s,j})^2}}
\end{equation}
This matrix reveals two types of dimensional correlations: diagonal elements $\mathbf{C}^{B_h}_{ii}$ measure the alignment of corresponding dimensions between ID and LLM embeddings, while off-diagonal elements $\mathbf{C}^{B_h}_{ij}~(i \neq j)$ measure redundancy across different dimensions.

To enhance correlation along corresponding dimensions while reducing redundancy across different dimensions, we define the alignment loss for the high-popularity item group as \cite{zbontar2021barlow}:
\begin{equation}
  \mathcal{L}^{B_h}_{FLA} = \underbrace{\sum_{i=1}^ {d} (1 - \mathbf{C}^{B_h}_{ii})^2}_{\text{invariance}} + \lambda \underbrace{\sum_{i=1}^{d} \sum_{j \neq i} (\mathbf{C}^{B_h}_{ij})^2}_{\text{redundancy reduction}}  
\end{equation}
where the invariance term promotes alignment within the corresponding dimensions, the redundancy reduction term promotes decoupling between different dimensions, and $\lambda$ controls the weight balance between the two terms. Similarly, we compute the feature-level alignment loss $\mathcal{L}^{B_l}_{FLA}$ for the low-popularity item group. The final feature-level alignment loss is the sum of the two losses:
\begin{equation}
    \mathcal{L}_{{FLA}} = \mathcal{L}^{B_h}_{{FLA}} + \mathcal{L}^{B_l}_{{FLA}}
\end{equation}

\noindent\textbf{Curriculum Learning Scheduler.}
Effective representation learning typically follows a hierarchical principle from simple to complex \cite{wang2021survey}, with item-level alignment and feature-level alignment precisely corresponding to this hierarchy: the former is a simple point-to-point task, while the latter is a complex alignment task over a space structure. If both constraints are imposed simultaneously during early training, premature space structure constraints may introduce noise, leading to suboptimal performance \cite{goyal2017accurate}. Therefore, inspired by curriculum learning \cite{graves2017automated,chen2021curriculum} we propose a cosine-scheduled to dynamically balance the two alignment objectives. Specifically, the weight of item-level alignment loss is defined as:
\begin{equation}
w_{i}(t) = (w_{\max} - w_{\min}) \cdot \frac{1 + \cos(2\pi t / T)}{2} + w_{\min}
\end{equation}
where $t$ denotes the training epoch, $T$ is the oscillation period, and $w_{\max}$ and $w_{\min}$ control the upper and lower bounds of the weight, respectively. Through this oscillation, the two alignment mechanisms complement each other with dynamically adjusted emphasis, enabling progressive refinement of item embeddings.

Integrating the item-level alignment loss $\mathcal{L}_{ILA}$ and feature-level alignment loss $\mathcal{L}_{FLA}$, the overall alignment loss is:
\begin{equation}
    \mathcal{L}_{align} = w_{i}(t) \cdot \mathcal{L}_{ILA} + (1 - w_{i}(t)) \cdot \mathcal{L}_{FLA}
\end{equation}

\subsection{Training and Inference}
\noindent\textbf{Training.} We adopt the commonly used binary cross-entropy (BCE) loss to train the sequential recommendation models \cite{kang2018self,dang2024augmenting,yue2024linear}. 
\begin{equation}
\begin{aligned}
\mathcal{L}_{rec} & = \operatorname{BCE}\left(H_u, E_u^{+}, E_u^{-}\right) \\ & = 
\left[\log \left(\sigma\left(H_u \cdot E_u^{+}\right)\right)+\log \left(1 -\sigma\left(H_u \cdot E_u^{-}\right)\right)\right]
\end{aligned}
\label{eq:rec}
\end{equation}
where $H_u, E_u^{+}$ and $E_u^{-}$ denote the representations of the user, positive and negative items, respectively. $\sigma(\cdot)$ is the sigmoid function. To jointly optimize recommendation performance and embedding alignment , we combine the recommendation loss with the alignment loss. The overall training objective is:
\begin{equation}
    \mathcal{L} = \mathcal{L}_{rec} + \alpha \cdot \mathcal{L}_{align}
\end{equation}
where $\alpha$ is a hyperparameter that balances the two objectives.

\vspace{0.3em}

\noindent\textbf{Inference.} During inference, the dual-level alignment has established structural consistency between the two embedding spaces and both embeddings are learned adequately. Therefore, to simplify the fusion process for efficiency, we use equal-weighted fusion of ID and LLM embeddings as final item embeddings for prediction.

\begin{table}[!t]
  \centering
  \caption{The statistics of three datasets.}
    \vspace{-1em}
    \begin{tabular}{ccccc}
    \toprule
    \textbf{Dataset} & \textbf{\# Users} & \textbf{\# Items} & \textbf{Sparsity} & \textbf{Avg.length} \\
    \midrule
    Beauty & 52,204 & 57,289 & 99.99\% & 7.56  \\
    Grocery & 32,126 & 39,264 & 99.98\% & 8.57  \\
    Yelp  & 15,720 & 11,383 & 99.89\% & 12.23  \\
    \bottomrule
    \end{tabular}%
  \label{tab:datasets}%
  \vspace{-1em}
\end{table}%

\begin{table*}[!t]
  \centering
  \caption{Performance comparison of different methods on Backbone models. The best performance is bolded. The `Impr.' represents the relative improvement of FAERec over the best baseline method.}
  \vspace{-1em}
  \renewcommand\arraystretch{1}
  \setlength{\tabcolsep}{1.5mm}
  \scalebox{0.97}{
    \begin{tabular}{cc|cccc|cccc|cccc}
    \toprule
    \multicolumn{2}{c|}{\multirow{3}[6]{*}{Backbone}} & \multicolumn{4}{c|}{Beauty}   & \multicolumn{4}{c|}{Grocery}     & \multicolumn{4}{c}{Yelp} \\
    \cmidrule{3-14}    \multicolumn{2}{c|}{} & \multicolumn{2}{c}{Overall} & \multicolumn{2}{c|}{Tail} & \multicolumn{2}{c}{Overall} & \multicolumn{2}{c|}{Tail} & \multicolumn{2}{c}{Overall} & \multicolumn{2}{c}{Tail} \\
    \cmidrule{3-14}    \multicolumn{2}{c|}{} & H@10  & N@10  & H@10  & \multicolumn{1}{c}{N@10} & H@10  & N@10  & H@10  & \multicolumn{1}{c}{N@10} & H@10  & N@10  & H@10  & N@10 \\
    \midrule
    \multirow{10}[1]{*}{SASRec} & Base  & 0.0331  & 0.0180  & 0.0113  & 0.0072  & 0.0332  & 0.0168  & 0.0086  & 0.0055  & 0.0240  & 0.0113  & 0.0060  & 0.0036  \\
          & CITIES & 0.0305  & 0.0159  & 0.0106  & 0.0064  & 0.0286  & 0.0144  & 0.0074  & 0.0045  & 0.0240  & 0.0111  & 0.0049  & 0.0024  \\
          & LOAM  & 0.0300  & 0.0158  & 0.0115  & 0.0076  & 0.0285  & 0.0148  & 0.0089  & 0.0059  & 0.0122  & 0.0055  & {0.0060}  & 0.0031  \\
          & MELT  & 0.0178  & 0.0088  & 0.0001  & 0.0000  & 0.0208  & 0.0101  & 0.0004  & 0.0003  & 0.0382  & 0.0188  & 0.0004  & 0.0002  \\
          & RLMRec-Con & \underline{0.0521}  & \underline{0.0287}  & \underline{0.0239}  & \underline{0.0132}  & \underline{0.0607}  & \underline{0.0344}  & \underline{0.0418}  & \underline{0.0248}  & 0.0406  & \underline{0.0212}  & \underline{0.0073}  & \underline{0.0041}  \\
          & RLMRec-Gen & 0.0439  & 0.0242  & 0.0122  & 0.0066  & 0.0471  & 0.0252  & 0.0119  & 0.0062  & 0.0391  & 0.0198  & 0.0051  & 0.0021  \\
          & LLMInit & 0.0377  & 0.0200  & 0.0103  & 0.0064  & 0.0434  & 0.0225  & 0.0087  & 0.0049  & 0.0374  & 0.0198  & 0.0044  & 0.0018  \\
          & LLM-ESR & 0.0269  & 0.0140  & 0.0049  & 0.0028  & 0.0298  & 0.0157  & 0.0051  & 0.0027  & \underline{0.0412}  & {0.0209}  & 0.0040  & 0.0019  \\
          & FAERec & \textbf{0.0642} & \textbf{0.0372} & \textbf{0.0633} & \textbf{0.0368} & \textbf{0.0776} & \textbf{0.0446} & \textbf{0.0718} & \textbf{0.0421} & \textbf{0.0461} & \textbf{0.0225} & \textbf{0.0090} & \textbf{0.0047} \\
          & Impr. & 23.22\% & 29.62\% & 164.85\% & 178.79\%  & 27.84\% & 29.65\% & 71.77\% & 69.76\% & 11.89\% & 6.13\% & 23.29\% & 14.63\% \\
    \midrule
    \multirow{10}[1]{*}{FMLP-Rec} & Base  & 0.0277  & 0.0153  & 0.0066  & 0.0047  & 0.0285  & 0.0146  & 0.0053  & 0.0037  & 0.0153  & 0.0072  & 0.0029  & 0.0014  \\
          & CITIES & 0.0238  & 0.0134  & 0.0062  & 0.0044  & 0.0252  & 0.0136  & 0.0048  & 0.0032  & 0.0153  & 0.0077  & 0.0015  & 0.0007  \\
          & LOAM  & 0.0225  & 0.0119  & 0.0057  & 0.0044  & 0.0221  & 0.0111  & 0.0048  & 0.0036  & 0.0115  & 0.0055  & 0.0022  & 0.0011  \\
          & MELT  & 0.0268  & 0.0143  & 0.0053  & 0.0030  & 0.0242  & 0.0122  & 0.0026  & 0.0014  & 0.0242  & 0.0119  & 0.0033  & 0.0013  \\
          & RLMRec-Con & 0.0458  & 0.0262  & \underline{0.0360}  & \underline{0.0217}  & \underline{0.0585}  & \underline{0.0322}  & \underline{0.0403}  & \underline{0.0229}  & 0.0232  & 0.0114  & \underline{0.0090}  & \underline{0.0048}  \\
          & RLMRec-Gen & \underline{0.0533}  & \underline{0.0305}  & 0.0269  & 0.0151  & 0.0550  & 0.0304  & 0.0285  & 0.0166  & 0.0178  & 0.0087  & 0.0026  & 0.0014  \\
          & LLMInit & 0.0383  & 0.0205  & 0.0096  & 0.0061  & 0.0406  & 0.0208  & 0.0091  & 0.0053  & 0.0265  & 0.0130  & 0.0038  & 0.0019  \\
          & LLM-ESR & 0.0224  & 0.0118  & 0.0051  & 0.0031  & 0.0274  & 0.0144  & 0.0081  & 0.0046  & \textbf{0.0308} & \textbf{0.0159} & 0.0040  & 0.0017  \\
          & FAERec & \textbf{0.0622} & \textbf{0.0361} & \textbf{0.0645} & \textbf{0.0379}  & \textbf{0.0749} & \textbf{0.0425} & \textbf{0.0737} & \textbf{0.0426} & \underline{0.0292}  & \underline{0.0149}  & \textbf{0.0097} & \textbf{0.0051} \\
          & Impr. & 16.70\% & 18.36\% & 79.17\% & 74.65\% & 28.03\% & 31.99\% & 82.88\% & 86.03\% & -5.19\% & -6.29\% & 7.78\% & 6.25\% \\
    \midrule
    \multirow{10}[2]{*}{LRURec} & Base  & 0.0349  & 0.0192  & 0.0108  & 0.0069  & 0.0351  & 0.0184  & 0.0080  & 0.0046  & 0.0288  & 0.0144  & 0.0033  & 0.0015  \\
          & CITIES & 0.0330  & 0.0175  & 0.0101  & 0.0062  & 0.0335  & 0.0175  & 0.0077  & 0.0047  & 0.0247  & 0.0115  & 0.0031  & 0.0013  \\
          & LOAM  & 0.0302  & 0.0160  & 0.0110  & \underline{0.0073}  & 0.0296  & 0.0150  & 0.0081  & 0.0050  & 0.0248  & 0.0119  & 0.0030  & 0.0015  \\
          & MELT  & 0.0242  & 0.0120  & 0.0018  & 0.0008  & 0.0251  & 0.0125  & 0.0008  & 0.0004  & 0.0387  & 0.0187  & 0.0004  & 0.0002  \\
          & RLMRec-Con & 0.0372  & 0.0205  & \underline{0.0117}  & \underline{0.0073}  & \underline{0.0602}  & \underline{0.0335}  & \underline{0.0239}  & \underline{0.0135}  & 0.0344  & 0.0172  & 0.0035  & 0.0020  \\
          & RLMRec-Gen & \underline{0.0418}  & \underline{0.0229}  & 0.0098  & 0.0058  & 0.0458  & 0.0249  & 0.0070  & 0.0041  & 0.0394  & 0.0202  & 0.0031  & 0.0013  \\
          & LLMInit & 0.0383  & 0.0205  & 0.0096  & 0.0061  & 0.0451  & 0.0234  & 0.0089  & 0.0051  & \underline{0.0415}  & \underline{0.0211}  & \underline{0.0057}  & \underline{0.0026}  \\
          & LLM-ESR & 0.0277  & 0.0144  & 0.0045  & 0.0024  & 0.0333  & 0.0173  & 0.0054  & 0.0030  & 0.0384  & 0.0194  & 0.0029  & 0.0013  \\
          & FAERec & \textbf{0.0644} & \textbf{0.0374} & \textbf{0.0617} & \textbf{0.0357} & \textbf{0.0759} & \textbf{0.0434} & \textbf{0.0675} & \textbf{0.0382} & \textbf{0.0433} & \textbf{0.0219} & \textbf{0.0086} & \textbf{0.0045} \\
          & Impr. & 54.07\% & 63.32\% & 427.35\% & 389.04\%  & 26.08\% & 29.55\% & 182.43\% & 182.96\% & 4.34\% & 3.79\% & 50.88\% & 73.08\% \\
    \bottomrule
    \end{tabular}
}%
  \label{tab:main_result}%
\end{table*}%

\subsection{Discussion}
Traditional approaches \cite{jang2020cities,kim2023melt} leverage collaborative signals from head items to improve tail-item performance. However, for items with extremely sparse interactions, the insufficient contextual signals limit the effectiveness of  such collaborative filtering-based knowledge transfer. Some works \cite{bao2023tallrec,gao2023chat} directly employ LLMs as recommenders, requiring LLM invocation at each inference step and introducing substantial computational overhead. Some works \cite{qu2024elephant,hu2024enhancing,ren2024representation} use LLM embeddings as auxiliary signals to enhance ID embeddings during training but discard semantic information at inference time. Some works \cite{liu2024llm,hu2025alphafuse} neglect structural consistency between embedding spaces when integrating the two embeddings. In contrast, our FAERec avoids high computational costs by using precomputed LLM embeddings. adaptive gating fusion leverages semantic knowledge to compensate for sparse collaborative signals in tail items. Dual-level alignment establishes structural consistency across embedding spaces, further improving the enhancement of LLM embeddings for recommendation. Moreover, our approach is model-agnostic and can be adapted to various SR models.

\section{Experiments}
\subsection{Experiment Setup}\label{Sec:setup}
\subsubsection{Datasets} The experiments are conducted on three datasets. Beauty and Grocery datasets are sourced from Amazon \cite{mcauley2015inferring} and correspond to the "Beauty" and "Grocery and Gourmet Food" categories, respectively. Yelp \footnote{\url{https://www.yelp.com/dataset}} contains user check-in histories along with their reviews. The maximum sequence length is set to 50 for all datasets. Table \ref{tab:datasets} presents the detailed statistics. We follow the preprocessing of the previous SR works \cite{kang2018self,kim2023melt,liu2024llm,dang2025data}. Since the sequential recommendation is commonly used with implicit feedback, we treat all rating records or reviews as interactions. Then, we exclude accounts with fewer than three interaction records, as the cold-start user problem is beyond our research scope.

\begin{table*}[!t]
  \centering
  \caption{The ablation study results on the Beauty dataset with three sequential recommendation backbones.}
  \vspace{-1em}
  \renewcommand\arraystretch{1}
  \setlength{\tabcolsep}{2.3mm}
  \scalebox{0.95}{
    \begin{tabular}{cc|cccc|cccc|cccc}
    \toprule
    \multicolumn{2}{c|}{\multirow{3}[6]{*}{Variant}} & \multicolumn{4}{c|}{SASRec}   & \multicolumn{4}{c|}{FMLP-Rec} & \multicolumn{4}{c}{LRURec} \\
\cmidrule{3-14}    \multicolumn{2}{c|}{} & \multicolumn{2}{c}{Overall} & \multicolumn{2}{c|}{Tail} & \multicolumn{2}{c}{Overall} & \multicolumn{2}{c|}{Tail} & \multicolumn{2}{c}{Overall} & \multicolumn{2}{c}{Tail} \\
\cmidrule{3-14}    \multicolumn{2}{c|}{} & H@10  & N@10  & H@10  & N@10  & H@10  & N@10  & H@10  & N@10  & H@10  & N@10  & H@10  & N@10 \\
    \midrule
    \multicolumn{2}{c|}{FAERec} & \textbf{0.0642} & \textbf{0.0372} & \textbf{0.0633} & \textbf{0.0368} & \underline{0.0622}  & \textbf{0.0361} & \textbf{0.0645} & \textbf{0.0379} & \textbf{0.0644} & \textbf{0.0374} & \textbf{0.0617} & \textbf{0.0357} \\
    \midrule
    \multicolumn{2}{c|}{w/o AGF} & 0.0534  & 0.0267  & 0.0359  & 0.0182  & \textbf{0.0624} & 0.0355  & 0.0516  & 0.0291  & 0.0561  & 0.0289  & 0.0369  & 0.0185  \\
    \multicolumn{2}{c|}{w/o ILA} & 0.0568  & 0.0318  & 0.0473  & 0.0260  & 0.0585  & 0.0343  & 0.0567  & 0.0331  & 0.0519  & 0.0289  & 0.0353  & 0.0192  \\
    \multicolumn{2}{c|}{w/o FLA} & 0.0626  & 0.0365  & \underline{0.0603}  & \underline{0.0355}  & 0.0597  & 0.0353  & 0.0630  & 0.0376  & 0.0632  & \underline{0.0369}  & \underline{0.0595}  & \underline{0.0346}  \\
    \multicolumn{2}{c|}{w/o CLS} & 0.0554  & 0.0309  & 0.0436  & 0.0239  & 0.0576  & 0.0330  & 0.0538  & 0.0305  & 0.0529  & 0.0292  & 0.0401  & 0.0218  \\
    \multicolumn{2}{c|}{w/o PG} & \underline{0.0636}  & \underline{0.0367}  & 0.0601  & 0.0350  & 0.0618  & \underline{0.0357}  & \underline{0.0642}  & \underline{0.0373}  & \underline{0.0636}  & \underline{0.0369}  & 0.0583  & 0.0339  \\
    \midrule
    \multicolumn{2}{c|}{Base} & 0.0331  & 0.0180  & 0.0113  & 0.0072  & 0.0277  & 0.0153  & 0.0066  & 0.0047  & 0.0349  & 0.0192  & 0.0108  & 0.0069  \\
    \bottomrule
    \end{tabular}%
 }
  \label{tab:ablation_results}%
  \vspace{-1em}
\end{table*}%

\begin{figure*}[!t]
  \centering
  \includegraphics[scale=0.35]{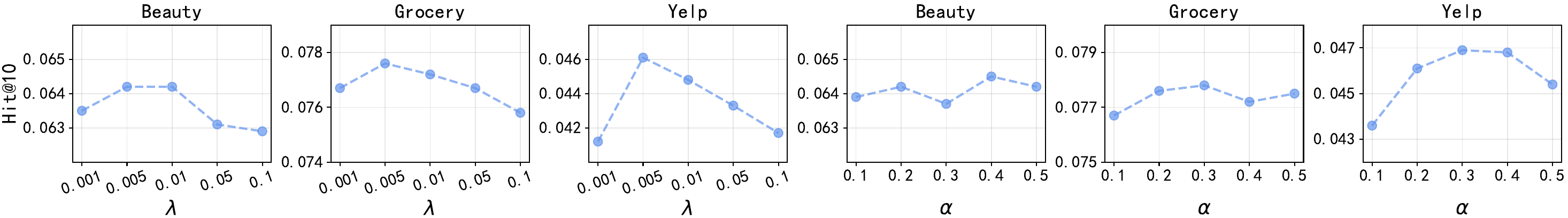}
  \vspace{-1em}
  \caption{Sensitivity analysis of hyperparameter $\lambda$ and $\alpha$ on three dataset with SASRec as the backbone.}
  \label{fig:hyperpara}
  \vspace{-0.5em}
\end{figure*}

\subsubsection{Backbones}
To verify the generalizability of our FAERec, we selected three representative general SR models as our backbones. 
\begin{itemize}
    \item SASRec \cite{kang2018self} leverages a unidirectional multi-head self-attention layers mechanism to model the casual transition correlations for the next item in the behavior sequence.
    \item FMLP-Rec \cite{zhou2022filter} introduces an all-MLP model with learnable filters for noise reduction in sequential recommendation.
    \item LRURec \cite{yue2024linear} proposes a linear recurrence model based on matrix diagonalization to capture transition patterns while improving training efficiency through parallel training.
\end{itemize}  

\subsubsection{Baselines}
We compare FAERec against various baseline methods, including three traditional long-tail models and three recent LLM-based methods.
\begin{itemize}
    \item CITIES \cite{jang2020cities} trains an embedding inference function using head item contexts to improve tail item quality.
    \item MELT \cite{kim2023melt} designs a mutually reinforcing dual-branch framework to tackle the long-tail user and item challenge. One branch transfers knowledge from head users to tail users, while the other transfers patterns from head items to tail items.
    \item LOAM \cite{yang2023loam} employs NicheWalk augmentation to capture session-level patterns with global context and Tail Session Mixup to synthesize interactions, enriching representations for tail items.
    \item RLMRec \cite{ren2024representation}, originally designed for collaborative filtering, is adapted to sequential recommendation in our work.It employs two knowledge transfer strategies to maximize mutual information between embeddings. RLMRec-Con maps LLM embeddings to collaborative space, while RLMRec-Gen 
    maps randomly masked ID embeddings to semantic space for reconstruction, both using similarity-based regularization.
    \item  LLMInit \cite{harte2023leveraging,hu2024enhancing,qu2024elephant} represents a category of methods that use language embeddings to initialize ID embeddings. Representative works include LLM2Bert4Rec \cite{harte2023leveraging}, SAID \cite{hu2024enhancing}, and Elephant \cite{qu2024elephant}, which initialize the item embedding layer with LLM embeddings and subsequently fine-tune them using interaction data.
    
    \item LLM-ESR \cite{liu2024llm} combines semantic knowledge from LLMs with collaborative signals for tail items through dual-view modeling, and introduce retrieval-augmented distillation for tail users.
\end{itemize}

\begin{figure*}[!t]
  \centering
  \includegraphics[scale=0.465]{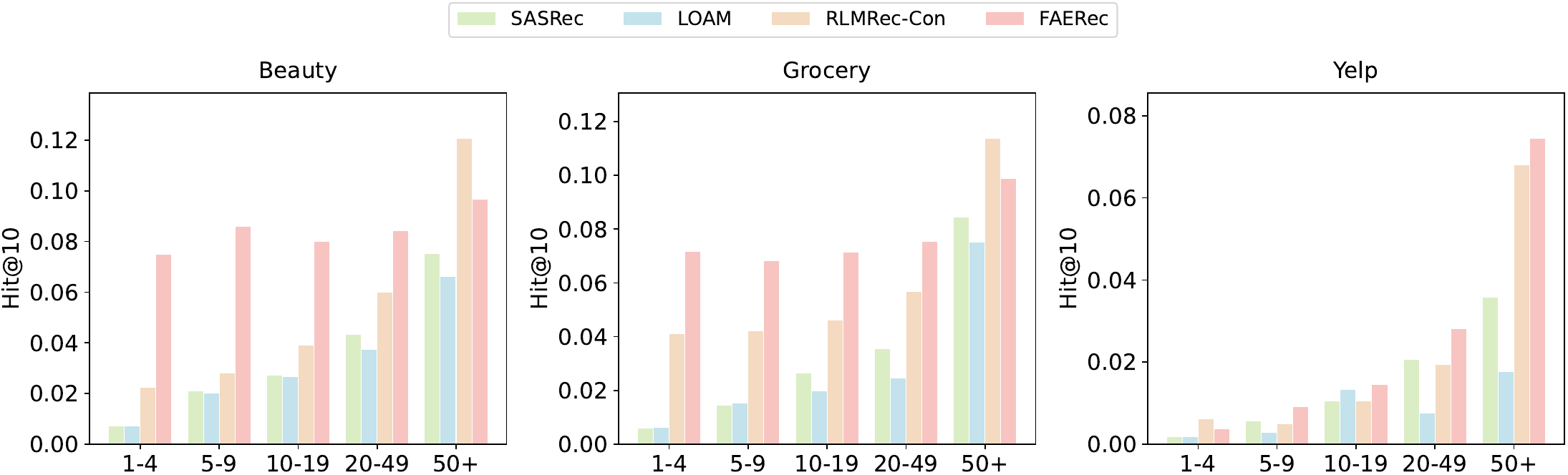}
  \vspace{-0.5em}
  \caption{The group analysis results on three dataset with SASRec as the backbone.}
  \label{fig:Group}
  \vspace{-0.5em}
\end{figure*}

\begin{table*}[!t]
  \centering
  \caption{Results on more sequential recommendation backbones.}
  \vspace{-1em}
  \renewcommand\arraystretch{1}
  \setlength{\tabcolsep}{2.15mm}
  \scalebox{0.95}{
    \begin{tabular}{c|cccc|cccc|cccc}
    \toprule
    \multirow{3}[5]{*}{Backbone} & \multicolumn{4}{c|}{Beauty}   & \multicolumn{4}{c|}{Grocery}  & \multicolumn{4}{c}{Yelp} \\
\cmidrule{2-13}          & \multicolumn{2}{c}{Overall} & \multicolumn{2}{c|}{Tail} & \multicolumn{2}{c}{Overall} & \multicolumn{2}{c|}{Tail} & \multicolumn{2}{c}{Overall} & \multicolumn{2}{c}{Tail} \\
\cmidrule{2-13}          & H@10  & N@10  & H@10  & N@10  & H@10  & N@10  & H@10  & N@10  & H@10  & N@10  & H@10  & N@10 \\
    \midrule
    NARM  & 0.0197  & 0.0097  & 0.0017  & 0.0010  & 0.0219  & 0.0114  & 0.0016  & 0.0008  & 0.0387  & 0.0195  & 0.0015  & 0.0009  \\
    w/FAERec & \textbf{0.0429} & \textbf{0.0226} & \textbf{0.0331} & \textbf{0.0176} & \textbf{0.0540} & \textbf{0.0286} & \textbf{0.0400} & \textbf{0.0213} & \textbf{0.0456} & \textbf{0.0223} & \textbf{0.0051} & \textbf{0.0021} \\
    LightSANs & 0.0252  & 0.0143  & 0.0059  & 0.0046  & 0.0291  & 0.0150  & 0.0052  & 0.0038  & 0.0167  & 0.0080  & 0.0024  & 0.0016  \\
    w/FAERec & \textbf{0.0597} & \textbf{0.0353} & \textbf{0.0614} & \textbf{0.0364} & \textbf{0.0714} & \textbf{0.0406} & \textbf{0.0727} & \textbf{0.0425} & \textbf{0.0254} & \textbf{0.0125} & \textbf{0.0115} & \textbf{0.0057} \\
    gMLP  & 0.0271  & 0.0163  & 0.0076  & 0.0058  & 0.0299  & 0.0157  & 0.0058  & 0.0040  & 0.0203  & 0.0093  & 0.0013  & 0.0007  \\
    w/FAERec & \textbf{0.0571} & \textbf{0.0325} & \textbf{0.0547} & \textbf{0.0311} & \textbf{0.0694} & \textbf{0.0383} & \textbf{0.0622} & \textbf{0.0340} & \textbf{0.0337} & \textbf{0.0169} & \textbf{0.0099} & \textbf{0.0048} \\
    \bottomrule
    \end{tabular}%
 }
  \label{tab:more_backbone}
  \vspace{-0.5em}
\end{table*}%

\subsubsection{Evaluation Settings} We adopt the leave-one-out strategy to partition sequences into training, validation, and test sets. Following previous practice \cite{krichene2020sampled,xie2022contrastive,yue2024linear,dang2023ticoserec,dang2023uniform}, we rank the prediction over the whole item set rather than negative sampling. The evaluation metrics include Hit Ratio@K (H@K) and Normalized Discounted Cumulative Gain@K (N@K). We report results with K $\in \{5, 10, 20\}$. 

\subsubsection{Implementation Details.} For all baselines, we adopt the implementation provided by the authors. For training, we set the batch size to 256 across all datasets and use the Adam \cite{2014Adam} optimizer with a learning rate of 0.001, $\beta_1=0.9$, and $\beta_2=0.999$. Following previous works \cite{liu2024llm,hu2025alphafuse}, we set the embedding size to 128. To ensure fair comparisons, we carefully set and tune all other hyperparameters of each method as reported and suggested in the original papers. 
For FAERec, we tune the $\lambda$, $\alpha$ in the range of  $\{0.001, 0.005, 0.01, 0.05, 0.1\}$, $\{0.1,0.2,0.3,0.4,0.5\}$, respectively. Furthermore, the embeddings of LLMs are derived from the API  named "text-embedding-ada-002" provided by OpenAI.

\subsection{Overall Comparison}
The experimental results of our FAERec and different baseline methods with various backbone SR models are presented in Table \ref{tab:main_result} \footnote{For MELT \cite{kim2023melt}, despite strictly following the authors' provided code and replicating the parameter settings and tuning ranges reported in the original paper, its performance consistently exhibits anomalies and underperforms in certain scenarios. We have observed similar phenomena in other published papers \cite{liu2024llm,li2025reembedding,li2025listwise}.}.

We have the following observations and conclusions:
\begin{itemize}
    \item Among the backbone models, LRURec achieves the best performance owing to its efficient linear recurrent modeling capability. For traditional long-tail methods, although they improve performance on tail items, their overall performance sometimes falls short of the base models due to the seesaw problem \cite{liu2024llm} (i.e., over-emphasizing tail items leads to degraded performance on head items). Notably, LLM-based methods consistently outperform traditional long-tail approaches, demonstrating the advantage of semantic information in handling extremely sparse data and the limitations of relying solely on collaborative signals.

    \item Among LLM-based methods, RLMRec-Con and RLMRec-Gen achieve leading overall performance in most cases, validating the effectiveness of cross-view knowledge transfer. RLMRec-Gen underperforms RLMRec-Con on tail items, as its masked reconstruction strategy further aggravates sparsity for tail items. In contrast, LLMInit and LLM-ESR perform relatively poorly. The former only initializes with semantic information and discards it during training, while the latter overlooks consistency of embedding spaces when combining ID and LLM embeddings.

    \item For our proposed FAERec, it achieves the best performance in nearly all cases. Notably, for tail items, the model achieves optimal performance across all cases. Through the adaptive fusion and dual-level alignment framework, FAERec effectively integrates collaborative signals and semantic information while maintaining consistency between embedding spaces. Even for scenarios with sparse interactions, FAERec leverages semantic information to compensate for the insufficiency of collaborative signals, thereby significantly enhancing the performance of tail items. 
\end{itemize}

\subsection{Ablation Study}
We conducted an ablation study to validate the effectiveness of various components within FAERec. We compared FAERec with the following variants:
1) w/o AGF: Removing adaptive gating fusion.
2) w/o ILA: Removing item-level alignment.
3) w/o FLA: Removing feature-level alignment.
4) w/o CLS: Setting both alignment losses to fixed weights without curriculum learning scheduler.
5) w/o PG: Removing popularity-based grouping in feature-level alignment.

As shown in Table \ref{tab:ablation_results}, all modules contribute to performance improvements. Removing adaptive gating fusion leads to significant performance degradation, particularly on tail items, highlighting the importance of incorporating semantic information into item embeddings for tail items. After removing either item-level or feature-level alignment, the models' performance metrics declined, demonstrating that both alignment strategies jointly facilitate model training. Notably, removing item-level alignment severely degrades tail-item performance, underscoring the need to establish basic item correspondences when combining heterogeneous embeddings. Performance decreases after removing the curriculum learning scheduler, even falling below that w/o FLA, verifying the importance of proper scheduling between the two alignments. If feature-level alignment is introduced prematurely, it injects noise during training and may even disrupt the correspondences established by item-level alignment. Additionally, popularity-based grouping prevents high-popularity items from dominating the alignment of low-popularity items at the feature level. Overall, these results validate that each component effectively addresses specific challenges for tail-item.

\subsection{Hyperparameter Investigation}
We further investigated the redundancy reduction weight $\lambda$ for feature-level alignment and the total alignment loss weight $\alpha$. The experimental results are shown in Figure \ref{fig:hyperpara}. As $\lambda$ increased from 0.001 to 0.1, the overall Hit@10 value of FAERec first rose and then declined, with the optimal value occurring in the range $\lambda \in \{0.005, 0.01\}$. This indicates that moderate redundancy reduction constraints can effectively maintain high alignment within corresponding dimensions while reducing information redundancy between different dimensional, thereby promoting consistency in the dimensional structures of the two embedding spaces.

For the alignment loss weight $\alpha$, we can observe that its optimal value typically lies within the range $\{0.3, 0.4\}$. When $\alpha$ is within this range, the alignment loss effectively establishes a stable and consistent corresponding between the ID and LLM embedding space, thereby promoting the effective integration of collaborative signals and semantic information. However, when $\alpha$ is too small, the alignment constraint fails to establish structural consistency between the two embedding spaces sufficiently. Conversely, when $\alpha$ is excessively large, overly stringent alignment can lead to convergence difficulties, causing the model to overemphasize space structure alignment at the expense of primary recommendation objectives, thereby degrading overall performance.These observations guide our hyperparameter configuration in practice.

\subsection{Group Analysis}
To investigate the long-tail problem more comprehensively, we divided items by their popularity into 5 groups.  As shown in Figure \ref{fig:Group}, we observe that FAERec can improve performance over the backbone across items of any popularity. In the 1-4 group, LOAM's performance on long-tail item groups lags significantly behind LLM-based models, demonstrating the limitations of traditional long-tail methods for extreme cold-start items. Compared with RLMRec-Con, our FAERec exhibits more performance improvements on long-tail items, especially for those in group 1-4 and group 5-9. However, FAERec slightly underperforms RLMRec-Con for popular items (e.g., 50+ group), indicating the inherent challenge of balancing performance between popular and long-tail items.

\subsection{Visualization}
\label{subsec:visualization}
\begin{figure}[!t]
  \centering
  \includegraphics[width=0.464\textwidth]{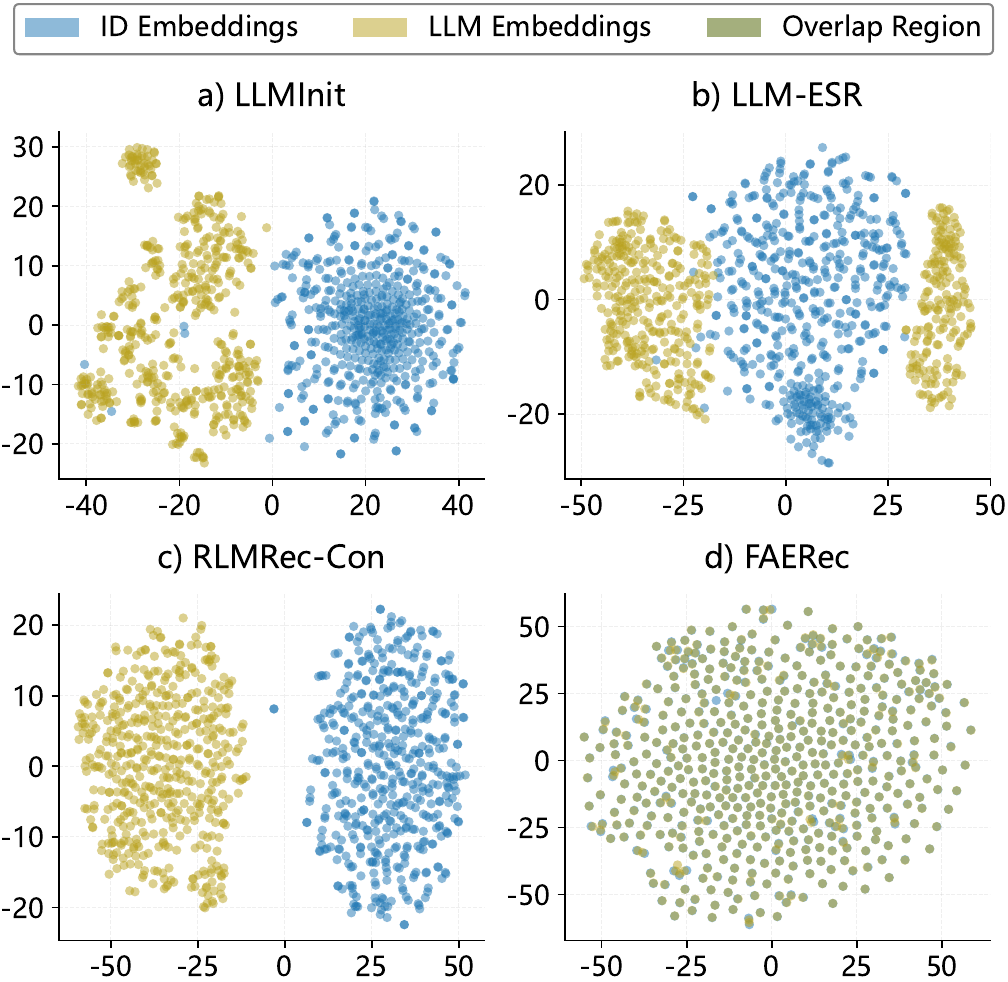}
  \hfill
  \vspace{-0.5em}
  \caption{Visualization of ID and LLM embedding spaces on the Beauty dataset with SASRec as the backbone.}
  \label{fig:tnse}
  \vspace{-0.5em}
\end{figure}

To verify the proposed structural inconsistency between the ID and LLM embedding spaces and the alignment effectiveness of FAERec, we visualize the embeddings of the same sampled item set using t-SNE in Figure \ref{fig:tnse}. We observe that LLMInit exhibits a scattered and uneven distribution since its LLM embeddings remain untrained. LLM-ESR lacks alignment strategies, resulting in significant discrepancies between ID and LLM embeddings. By introducing cross-view contrastive learning, RLMRec-Con achieves similar distribution patterns across ID and LLM embeddings. In contrast, FAERec goes further: through the dual-level alignment framework, it not only produces more uniform distributions but also establishes high structural consistency across the two embedding spaces, thereby improving the quality of item embeddings.

\subsection{Generalizing to More Backbones}
To further verify the generalization capability of FAERec, we apply it to more SR backbones with different architectures:
\begin{itemize}
    \item NARM \cite{li2017neural}: An RNN-based model that integrates attention mechanisms into recurrent neural networks to capture both global sequential behaviors and local key intentions.
    \item LightSANs \cite{fan2021lighter}: A lightweight self-attention model that projects user interests through low-rank factorized self-attention and generates context-aware item representations.
    \item gMLP \cite{liu2021pay}: An MLP-based model that replaces self-attention with spatial gating units, capturing sequential dependencies through cross-position multi-layer perceptrons.
\end{itemize}
As shown in Table \ref{tab:more_backbone}, we observe that FAERec achieves significant improvements across all backbone models consistently. For overall performance, FAERec achieves average improvements of 124.1\%, 148.3\%, and 48.0\% across the three datasets, respectively. For long-tail item recommendation, the improvements are even more substantial, reaching 1032.5\%, 1500.2\%, and 376.0\%, respectively. These results validate the effectiveness and generalization capability of our method, particularly demonstrating outstanding performance in long-tail item recommendation scenarios.

\subsection{Diversity Analysis}
Diverse recommendations deliver surprise and discovery value, enhancing long-term user satisfaction \cite{kim2019sequential}. Therefore, we further evaluate the diversity of FAERec's recommendation. Following previous research \cite{yang2023loam,liu2020long}, we adopt Coverage@K (Cov@K) and Tail\_Coverage@K (TCov@K), which measure the number of distinct items appearing in the top-K recommendations and the number of long-tail items, respectively. Their definitions are as follows:
\begin{equation}
\begin{aligned}
\text{Coverage@K} &= \frac{\left| \bigcup_{u \in \mathcal{U}} L_K(u) \right|}{|\mathcal{V}|} \\
\text{Tail\_Coverage@K} &= \frac{\left| \bigcup_{u \in \mathcal{U}} L_K^{\text{tail}}(u) \right|}{|\mathcal{V}_{\text{tail}}|}
\end{aligned}
\end{equation}
where $L_K(u)$ denotes the top-K recommendation list generated for user $u$. Specifically, $L_K^{\text{tail}}(u)$ represents the subset of $L_K(u)$ that contains items belonging to the long-tail item set $\mathcal{V}_{\text{tail}}$.

As shown in Table \ref{tab:diversity}, although LOAM and RLMRec-Con improve the prediction accuracy for long-tail items, their contribution to recommendation diversity remains limited. In contrast, our proposed FAERec not only achieves improvements in tail-item performance but also significantly enhances recommendation diversity by effectively integrating collaborative signals with semantic information to compensate for the inherent disadvantages of long-tail items.

\begin{table}[!t]
  \centering
  \caption{Diversity comparison of different methods on the Beauty dataset with three SR backbones.}
  \vspace{-1em}
  \renewcommand\arraystretch{1}
  \setlength{\tabcolsep}{1.5mm}
  \scalebox{0.75}{
    \begin{tabular}{c|cc|cc|cc}
    \toprule
    \multirow{2}[5]{*}{Method} & \multicolumn{2}{c|}{SASRec} & \multicolumn{2}{c|}{FMLP-Rec} & \multicolumn{2}{c}{LRURec} \\
    \cmidrule{2-7}          & Cov@10 & TCov@10 & Cov@10 & TCov@10 & Cov@10 & TCov@10 \\
    \midrule
    Base  & \underline{0.6972}  & \underline{0.6266}  & 0.5273  & 0.4267  & 0.6679  & 0.5920  \\
    LOAM  & 0.6419  & 0.5634  & 0.5092  & 0.4057  & \underline{0.6768}  & \underline{0.6036}  \\
    RLMRec-Con & 0.5019  & 0.4141  & \underline{0.7010}  & \underline{0.6442}  & 0.5477  & 0.4554  \\
    LLM-ESR & 0.3513  & 0.2251  & 0.5176  & 0.4215  & 0.4075  & 0.2831  \\
    FAERec & \textbf{0.8538} & \textbf{0.8249} & \textbf{0.8571} & \textbf{0.8298} & \textbf{0.8612} & \textbf{0.8315} \\
    \bottomrule
    \end{tabular}
}
  \label{tab:diversity}
  \vspace{-1em}
\end{table}

\section{CONCLUSION}
This paper introduces FAERec, a fusion and alignment enhancement framework with LLMs tailored for tail item sequential recommendations. To tackle the challenges of long-tail items, FAERec improves item embedding quality and enhances long-tail recommendation performance by effectively integrating collaborative signals with semantic information. Specifically, we design an adaptive gated fusion mechanism to capture fine-grained complementary patterns between the two embeddings. To further enhance fusion effectiveness, we introduce a dual-layer alignment framework, item-level alignment and feature-level alignment, establishing point-to-point correspondences and spatial structural consistency between the two embedding spaces. Additionally, a curriculum learning scheduler balances the two alignment objectives to ensure training stability. We conducted experiments on three real-world datasets to comprehensively validate the effectiveness and generalizability of FAERec. 

\section{Acknowledgments}
This research is partially supported by the National Natural Science Foundation of China under Grant No. 62576083. This research is also supported by the Ministry of Education, Singapore, under its Academic Research Fund (AcRF) Tier 1 grant, and funded through the SMU-SUTD Internal Research Grant Call (SMU-SUTD 2023\_02\_01), and in part by the Ministry of Education, Singapore, under its Academic Research Fund Tier 2 (Award No. MOE-T2EP201230015). The authors greatly appreciate the anonymous reviewers for their valuable comments.



\bibliographystyle{ACM-Reference-Format}
\bibliography{references}


\end{document}